# BISTATIC SCATTERING FORWARD MODEL VALIDATION USING GNSS-R OBSERVATIONS

*Amir Azemati[(1)], Mahta Moghaddam[(1)], Arvind Bhat[(2)]*

(1) Ming Hsieh Department of Electrical Engineering
University of Southern California, Los Angeles, CA, 90089
(2) Intelligent Automation INC. (IAI), Rockville, MD, 20855

## ABSTRACT

In this paper we advance a previously developed bistatic scattering forward model to include the circularly polarized incident and scattered waves, which is the case for Global Navigation Satellite System (GNSS) reflectometry. This model development enables retrieval of soil moisture from Signals of Opportunity (SoOp) bistatic observations, e.g., from the Cyclone Global Navigation Satellite System (CYGNSS) observations and GNSS Reflectometer Instrument for Bistatic Synthetic Aperture Radar (GRIBSAR). In order to validate the forward model with measured data, we present a method to construct the Delay Doppler Map (DDM) from simulations of the forward model. The forward model Radar Cross Section (RCS) predictions will be compared with actual measurements. The validated model is intended for use in soil moisture retrievals.

## 1. INTRODUCTION

Global land surface observations of soil moisture are essential for better understanding of how water, carbon, and energy cycles are linked. Soil moisture mapping has crucial impacts on many areas of human interest including weather and climate forecasting, wildfire and flood prediction, agriculture and drought analysis, and human health [1-5]. Soil moisture can be retrieved using active or passive microwave observations. Active microwave systems, or radars, can provide higher resolution estimates of soil moisture than their passive counterparts. Conventional monostatic radars, however, are expensive, and it is a total loss if their transmitter fails, as in the case of the Soil Moisture Active Passive (SMAP) mission. In contrast, bistatic radars are much more resilient in that their transmitters are pervasively present, such as in the case of GPS/GNSS transmitters. Therefore, it is important to advance novel soil moisture retrieval techniques by using the available SoOp from these sources [6-12]. In our previous works [3-5], a fully bistatic scattering forward model for vegetated terrain has been developed, which provides the possibility to exploit SoOp for soil moisture retrieval over various land covers including forests. The proposed bistatic scattering model has three main scattering mechanisms: direct ground scattering, vegetation volume scattering, and double bounce scattering from ground and vegetation layer. It also includes the Right-Hand Circularly Polarized (RHCP) incident and linearly polarized scattered wave cases [3-5]. The RHCP GNSS incident waves produce both RHCP ($\sigma_{RR}$) and Left-Hand Circularly Polarized (LHCP) scattered waves ($\sigma_{LR}$) upon reflection from the ground surface. Thus, in section 2 of this paper we extend our proposed bistatic scattering model to include the RHCP incident and circularly polarized (RHCP and LHCP) scattered waves scenarios. In section 3 of this paper we present a method for constructing DDMs by using the forward bistatic scattering model. A validation technique for the forward model is presented in section 4, in which the developed bistatic scattering model will be used to generate the DDMs (e.g. Fig. 1) for Tonzi ranch area in California. The accuracy of the bistatic model will be assessed with data available from bistatic platforms such as from CYGNSS and GRIBSAR [2-4].

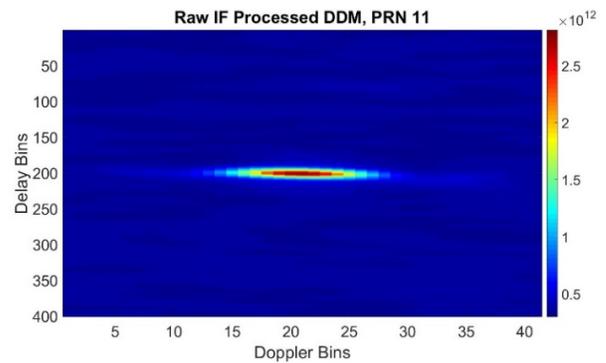

**Fig. 1.** An example of overland (Herbaceous Amazon forest) Delay Doppler Map (DDM), which is derived from CYGNSS raw IF data. The raw IF processed data are measured in "counts" unit, which is proportional to the total received signal power. The center bin of this DDM represents scattering in specular direction, which has the highest Signal to Noise Ratio (SNR).

## 2. CIRCULAR/CIRCULAR BISTATIC MODEL

The current version of bistatic scattering model includes the linearly polarized incident and scattered wave cases ($\sigma_{HH}, \sigma_{VV}, \sigma_{HV}, \sigma_{VH}$) [4]. However, in order to utilize the GNSS signals as the sources of opportunity, the forward bistatic scattering model needs to handle circularly polarized incident and scattered wave scenarios ($\sigma_{RR}, \sigma_{LR}$). The circularly polarized received signal scattered in specular direction can be derived from their corresponding reflection coefficients ($\Gamma_{RR}, \Gamma_{LR}$). The relationship between circularly polarized reflection coefficients ($\Gamma_{RR}, \Gamma_{LR}$) and circularly polarized specular radar cross sections ($\sigma_{RR}, \sigma_{LR}$) can be expressed as [13]:

$$\sigma_{RR} = 4\pi |\Gamma_{RR}|^2 \quad (1)$$

$$\sigma_{LR} = 4\pi |\Gamma_{LR}|^2 \quad (2)$$

Moreover, as shown in the compact polarimetry study of [13-14], the circularly polarized reflection coefficients can be written in terms of the linearly polarized reflection coefficients ($\Gamma_{HH}, \Gamma_{VV}$) as follows:

$$\Gamma_{RR} = -\frac{(\Gamma_{hh}+\Gamma_{vv})}{2} \quad (3)$$

$$\Gamma_{LR} = -\frac{(\Gamma_{vv}-\Gamma_{hh})}{2} \quad (4)$$

Equations (1)-(4) can be utilized to express the circularly polarized RCS ($\sigma_{RR}, \sigma_{LR}$) in terms of the linearly polarized RCS ($\sigma_{HH}, \sigma_{VV}$):

$$\sigma_{RR} = \frac{\sigma_{HH}+\sigma_{VV}}{4} + \sqrt{\sigma_{HH}\sigma_{VV}} \quad (5)$$

$$\sigma_{LR} = \frac{\sigma_{HH}+\sigma_{VV}}{4} - \sqrt{\sigma_{HH}\sigma_{VV}} \quad (6)$$

Therefore, in order to compute the $\sigma_{RR}$ and $\sigma_{LR}$, first our forward bistatic scattering model will be used to predict the radar cross section values corresponding to the linearly polarized received signal ($\sigma_{HH}, \sigma_{VV}$), and then equations (5) and (6) will be utilized to derive the cross section values related to the circularly polarized incident and scattered waves ($\sigma_{RR}, \sigma_{LR}$). For the case of scattering in the non-specular direction we will follow the same approach described above, however, we need to use the absolute value of co-polarization scattering elements of scattering matrix in lieu of reflection coefficients, namely ($S_{HH}, S_{VV}$) [4-5].

## 3. CONSTRUCTING DDM BY USING THE ADVANCED BISTATIC MODEL

The measurements taken from SoOp reflections from the Earth surface are represented as Delay Doppler Maps (DDMs) (Fig. 1) [2-3]. It is shown in section 2 that it is possible to use a polarimetric bistatic scattering model to compute the relationship between circularly and linearly polarized radar cross sections ($\sigma_{RR}, \sigma_{LR}$) and ($\sigma_{HH}, \sigma_{VV}, \sigma_{HV}, \sigma_{VH}$). The next step is to validate these RCS values predicted by the forward bistatic scattering model with the actual measurements of CYGNSS and GRIBSAR receivers. Since the CYGNSS and GRIBSAR data are presented as DDMs, the proposed bistatic scattering model needs to be utilized to generate these DDMs for validation purposes. In addition, the DDMs' unit will be converted from Watts to normalized circularly polarized RCS in order to be consistent with the forward model. Therefore, the conversion method, which is described in [3], will be applied to convert the scattered signal power DDM in Watts (Fig. 2) to normalized bistatic scattering RCS.

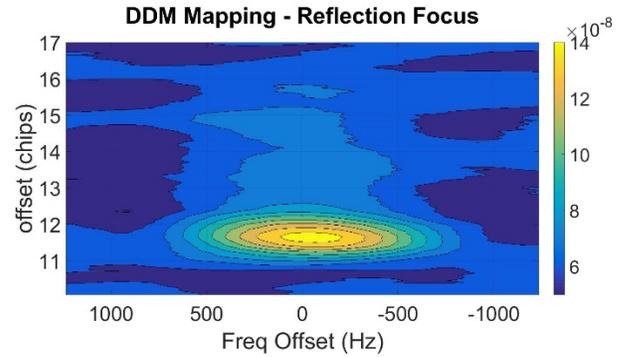

**Fig. 2.** DDM of the Tonzi ranch area in California, which demonstrates the GPS scattered signal power measured by the GRIBSAR.

According to Fig. 2, the bright yellow bin at zero Doppler, which has the highest SNR, corresponds to the specular bin and the area around this bin is called the glistening zone. Moreover, the size of the glistening zone is determined by the root mean square of surface roughness distribution [2]. In order to construct the over-land DDM with the proposed bistatic scattering model, we first compute the specular bin RCS [3] and set the specular bin as the reference point. Each bin around the specular direction has a specific set of scattering ($\theta_s$) and azimuth angles ($\varphi_s$). Depending on the size of the glistening zone, which is proportional to the random distribution of surface roughness [2], and the receiver distance from that region, it is possible to set separate ranges for scattering ($\theta_s$) and azimuth ($\varphi_s$) angles. The specular direction is located at the center of these ranges and the size of the glistening zone (area of interest) [2] specifies the lower and upper bounds of azimuth and scattering angle ranges. After setting the ranges for scattering and azimuth angles, we construct the over-land DDM by running the forward bistatic scattering model for each of the DDM bins, including scattering contributions from their corresponding scattering and azimuth angles. For terrains with small surface roughness a combination of Small Perturbation Method (SPM) and Kirchoff approximation (KA) is used to compute the bistatic scattering from glistening zone and specular direction (bin), respectively [15-18]. However, for terrains with high surface roughness, SPM + KA is no longer acceptable and the Stabilized Extended Boundary Condition Method (SEBCM) [19] will be applied to compute the ground bistatic scattering contribution in the future. Unlike the analytical methods (SPM+KA), the SEBCM calculates the coherent (specular direction) and the non-coherent (glistening-zone) scattering RCS all in one shot at the cost of higher computational time.

## 4. VALIDATION OF BISTATIC MODEL

In the previous section we discussed the method used to construct the over-land DDMs with the developed forward scattering model. The next step after constructing the DDMs is the forward model validation with real measured data available from sources such as CYGNSS and GRIBSAR, using in-situ data from the Soil moisture Sensing Controller and oPtimal Estimator (SoilSCAPE) network [20] to parameterize the forward model. In order to validate the forward model, the measured DDMs will be compared with the DDMs simulated by utilizing our forward bistatic scattering model and the method described in Section 3. In August 2018, Intelligent Automation Inc. (IAI) flew the GRIBSAR instrument onboard an airplane in order to measure bistatic GPS signals scattered from Tonzi-ranch in California (Fig. 3). This site is home to one of the SoilSCAPE in-situ sensor networks, providing several nodes for soil moisture profile observations at sub-hourly temporal resolution. Thus, it provides an excellent opportunity to validate the bistatic scattering model predictions with actual GNSS-R data from GRIBSAR.

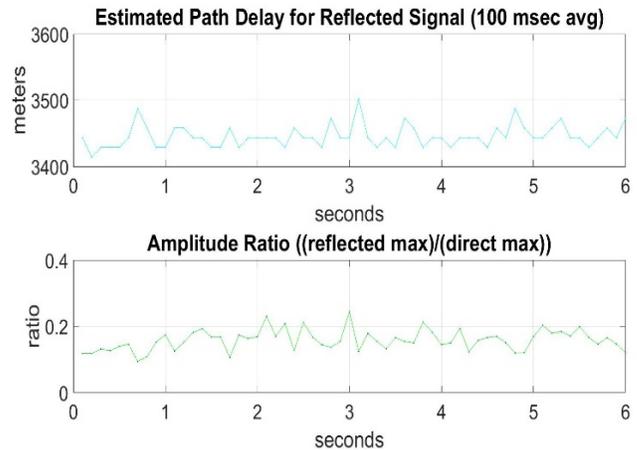

**Fig. 3.** Estimated path delay of reflected GPS signal and the amplitude ratio of reflected and direct GPS signal, which are measured by the GRIBSAR for the Tonzi ranch region in California.

## 5. RESULTS AND SUMMARY

To validate the forward model and the associated DDMs from Tonzi ranch, we first simulate the DDM. To do so, we use the field-measured values of soil moisture profiles from the SoilSCAPE network and our existing database of surface roughness and vegetation parameters that we have previously measured in the field, to parametrize the radar scattering model. We then compute the circularly polarized bistatic scattering cross sections corresponding to each DDM bin according to their scattering and azimuth angles, followed by filling in each DDM bin with the points on the ground that contribute to that bin. We compare the DDM thus simulated with the observed DDM, after making proper adjustments to measurement units and applying the required calibration parameters. Results of this analysis and comparison will be shown at the presentation. In the future, we will use the calibrated data from GRIBSAR, proposed validated forward model, and well-tested inverse scattering algorithms [21] to retrieve soil moisture at Tonzi ranch and other locations where data might be available. Once demonstrated, the method will be used more broadly for soil moisture retrievals from CYGNSS data.

# 6. REFERENCES

[1] D. Entekhabi *et al.*, "The Soil Moisture Active Passive (SMAP) Mission," in *Proceedings of the IEEE*, vol. 98, no. 5, pp. 704-716, May 2010.

[2] Cyclone Global Navigation Satellite System (CYGNSS) Handbook, University of Michigan, Ann Arbor, MI, 2016.

[3] A. Azemati, M. Moghaddam and A. Bhat, "Relationship Between Bistatic Radar Scattering Cross Sections and GPS Reflectometry Delay-Doppler Maps Over Vegetated Land in Support of Soil Moisture Retrieval," *IGARSS 2018 - 2018 IEEE International Geoscience and Remote Sensing Symposium*, Valencia, 2018, pp. 7480-7482. doi: 10.1109/IGARSS.2018.8517345

[4] A. Azemati and M. Moghaddam, "Circular-linear polarization signatures in bistatic scattering models applied to signals of opportunity," *2017 International Conference on Electromagnetics in Advanced Applications (ICEAA)*, Verona, 2017, pp. 1825-1827.

[5] A. Azemati and M. Moghaddam, "Modeling and analysis of bistatic scattering from forests in support of soil moisture retrieval," *2017 IEEE International Symposium on Antennas and Propagation & USNC/URSI National Radio Science Meeting*, San Diego, CA, 2017, pp. 1833-1834.

[6] N. Pierdicca, L. Pulvirenti, F. Ticconi and M. Brogioni, "Radar Bistatic Configurations for Soil Moisture Retrieval: A Simulation Study," in *IEEE Transactions on Geoscience and Remote Sensing*, vol. 46, no. 10, pp. 3252-3264, Oct. 2008.

[7] N. Rodriguez-Alvarez *et al.*, "Soil Moisture Retrieval Using GNSS-R Techniques: Experimental Results Over a Bare Soil Field," in *IEEE Transactions on Geoscience and Remote Sensing*, vol. 47, no. 11, pp. 3616-3624, Nov. 2009.

[8] A. Camps *et al.*, "Sensitivity of GNSS-R Spaceborne Observations to Soil Moisture and Vegetation," in *IEEE Journal of Selected Topics in Applied Earth Observations and Remote Sensing*, vol. 9, no. 10, pp. 4730-4742, Oct. 2016.

[9] R. Shah, C. Zuffada, C. Chew, M. Lavalle, X. Xu and A. Azemati, "Modeling bistatic scattering signatures from sources of opportunity in P-Ka bands," *2017 International Conference on Electromagnetics in Advanced Applications (ICEAA)*, Verona, 2017, pp. 1684-1687.

[10] C. C. Chew, E. E. Small, K. M. Larson and V. U. Zavorotny, "Effects of Near-Surface Soil Moisture on GPS SNR Data: Development of a Retrieval Algorithm for Soil Moisture," in *IEEE Transactions on Geoscience and Remote Sensing*, vol. 52, no. 1, pp. 537-543, Jan. 2014.

[11] K. M. Larson, J. J. Braun, E. E. Small, V. U. Zavorotny, E. D. Gutmann and A. L. Bilich, "GPS Multipath and Its Relation to Near-Surface Soil Moisture Content," in *IEEE Journal of Selected Topics in Applied Earth Observations and Remote Sensing*, vol. 3, no. 1, pp. 91-99, March 2010.

[12] N. Rodriguez-Alvarez *et al.*, "Land Geophysical Parameters Retrieval Using the Interference Pattern GNSS-R Technique," in *IEEE Transactions on Geoscience and Remote Sensing*, vol. 49, no. 1, pp. 71-84, Jan. 2011.

[13] Ulaby, F.T., Long, D.G., Blackwell, W.J., Elachi, C., Fung, A.K., Ruf, C., Sarabandi, K., Zebker, H.A. and Van Zyl, J., 2014. *Microwave radar and radiometric remote sensing* (Vol. 4, No. 5, p. 6). Ann Arbor: University of Michigan Press.

[14] J. D. Ouellette *et al.*, "A Simulation Study of Compact Polarimetry for Radar Retrieval of Soil Moisture," in *IEEE Transactions on Geoscience and Remote Sensing*, vol. 52, no. 9, pp. 5966-5973, Sept. 2014.

[15] Ishimaru, Akira. *Wave propagation and scattering in random media*. Vol. 2. New York: Academic press, 1978.

[16] L. Tsang *et al.*, "Active and Passive Vegetated Surface Models With Rough Surface Boundary Conditions From NMM3D," in *IEEE Journal of Selected Topics in Applied Earth Observations and Remote Sensing*, vol. 6, no. 3, pp. 1698-1709, June 2013.

[17] A. Tabatabaeenejad and M. Moghaddam, "Bistatic scattering from three-dimensional layered rough surfaces," in *IEEE Transactions on Geoscience and Remote Sensing*, vol. 44, no. 8, pp. 2102-2114, Aug. 2006.

[18] Tsang, Leung, et al. *Scattering of electromagnetic waves, numerical simulations*. Vol. 25. John Wiley & Sons, 2004.

[19] X. Duan and M. Moghaddam, "Stabilized extended boundary condition method for 3D electromagnetic scattering from arbitrary random rough surfaces," *2010 IEEE Antennas and Propagation Society International Symposium*, Toronto, ON, 2010, pp. 1-4.

[20] M. Moghaddam *et al.*, "A Wireless Soil Moisture Smart Sensor Web Using Physics-Based Optimal Control: Concept and Initial Demonstrations," in *IEEE Journal of Selected Topics in Applied Earth Observations and Remote Sensing*, vol. 3, no. 4, pp.522-535, Dec.2010. doi: 10.1109/JSTARS.2010.2052918

[21] Tabatabaeenejad, A., and M. Moghaddam, "Inversion of subsurface properties of layered dielectric structures with random rough interfaces using the method of simulated annealing," *IEEE Trans. Geosci. Remote Sensing*, 47(7), 2035-2046, 2009